\documentclass[prd,a4paper,twocolumn,preprintnumbers,floatfix,aps,nofootinbib]{revtex4}

\usepackage{latexsym}
\usepackage{epsfig}
\usepackage{amssymb}

\newcommand{\bea}{\begin{eqnarray}}
\newcommand{\eea}{\end{eqnarray}}
\newcommand{\be}{\begin{equation}}
\newcommand{\ee}{\end{equation}}

\begin{document}

\title{Viable Palatini-f(R) cosmologies with generalized dark matter}
\date{\today}

\author{Tomi Koivisto}
\email{tomikoiv@pcu.helsinki.fi}
\affiliation{Helsinki Institute of Physics, P.O. Box 64, FIN-00014 Helsinki, 
Finland}
\affiliation{Department of Physical Sciences, Helsinki University, P.O. 
Box 64, FIN-00014 Helsinki, Finland}

\begin{abstract}

We study the formation of large-scale structure in universes dominated by dark
matter and driven to accelerated expansion by f(R) gravity in the Palatini
formalism. If the dark matter is cold, practically all of these models are ruled out
because they fail to reproduce the observed matter power spectrum.
We point out that if the assumption that dark matter is perfect and pressureless
at all scales is relaxed, nontrivial alternatives to a cosmological
constant become viable within this class of modified gravity models.

\end{abstract}

\maketitle
 
\section{Introduction}

The $f(R)$ theories of gravity have been intensively explored in recent years
as possible alternatives to dark energy\cite{Nojiri:2006ri}.
This class of extensions of general relativity 
is defined via a seemingly simple generalization of the Einstein-Hilbert action by 
allowing nonlinear interactions of the Ricci scalar as  
\be \label{action}
 S = 
  \frac{1}{16\pi G}
  \int d^4x \sqrt{-g}f\left( g^{\mu\nu}R_{\mu\nu}(\Gamma ) \right) + S_m,
 \ee
where $S_m$ is the matter action. When
$\Gamma$ is taken to be the metric-compatible (Christoffel) connection,
the action (\ref{action}) represents the so called metric $f(R)$ 
models \cite{Carroll:2003wy,Nojiri:2003rz,Nojiri:2003ft,Faraoni:2006fx,Faraoni:2005ie}. Though
a realistic universe expansion history may be reconstructed from these models 
\cite{Nojiri:2006gh,Capozziello:2006dj}, they
seem to be ruled out as alternatives to dark energy because of their 
consequences to the Solar system physics \cite{Chiba:2003ir,Erickcek:2006vf,Kainulainen:2007bt}, see 
however \cite{Hu:2007nk}. If the connection $\Gamma$ is promoted to an independent variable, the action
(\ref{action}) represents the $f(R)$ models in the so called Palatini 
formalism \cite{Ferraris:1992dx,Vollick:2003aw,Flanagan:2003iw,Meng:2003en,Allemandi:2005qs,Poplawski:2006kv},
which instead appear to pass the Solar system tests \cite{Kainulainen:2006wz}. 
The viability of the small-scale limit of these Palatini-$f(R)$ models has been an 
issue of debate\cite{Flanagan:2003rb,Vollick:2003ic}. A recent claim of 
violations of the equivalence principle \cite{Olmo:2006zu}, though partly erronous \cite{Koivisto:2005yk}, has resurfaced doubts about the physical tenability of
the Palatini-$f(R)$ gravity \cite{Olmo:2006zu,Barausse:2007pn}. These 
have been already considerably clarified by Kainulainen {\it et al} \cite{Kainulainen:2007bt},
and the problems which perhaps remain at tiny scales could also dissipate if one 
entertains these models as a macroscopic limit of a possibly more fundamental 
description of spacetime with the aid of truly metric-affine degrees 
of freedom \cite{Sotiriou:2006qn}.
 
This suggests it worthwhile to seriously reconsider the cosmology of the Palatini-$f(R)$ 
models. Expansionwise, they can generate a viable sequence of radiation dominated, matter dominated and 
accelerating era matching with the constraints, as has been shown for various 
parameterizations of the function $f(R)$, most often with some power-law forms 
(with one, two or three powers of $R$), but also with square-root, logarithmic and exponential 
forms for the curvature correction 
terms \cite{Capozziello:2004vh,Amarzguioui:2005zq, Borowiec:2006hk,Movahed:2007cs,Fay:2007gg}. 
However, the inhomogeneous evolution present in any realistic universe has not been 
succesfully reconciled with observations in these models. 
At the background level the additional derivative terms in the field equations
due to nonlinearity in $f(R)$ can play the role of an effective smooth dark energy. 
Meanwhile the spatial gradients of these extra terms cannot be neglected for 
cosmological perturbations \cite{Koivisto:2005yc} which then assume unusual behaviour that 
is at odds with the observed distribution of galaxies \cite{Koivisto:2006ie} and with the
cosmic microwave background spectrum \cite{Li:2006vi}, even if the nonlinear part of the action 
is exponentially suppressed at late times \cite{Li:2006ag}. In quantitative terms,
parameterizing the $f(R)- R \sim R^\beta$, data analysis constrains  $|\beta| < 10^{-5}$
\cite{Koivisto:2005yc,Li:2006vi}. 
Thus, while these models are allowed by local experiments, 
they are extremely tightly constrained by cosmological data.

In this paper our purpose is to investigate under which conditions these
cosmological constraints on the Palatini-$f(R)$ could be loosened. 
Our approach is to allow generalized dark matter (GDM) with possible (isotropic or anisotropic) 
pressures in the matter sector. The gravity sector we assume to be given by a general nonlinear function $f(R)$ 
appearing in Eq.(\ref{action}), when understood in the Palatini formalism which yields second order field 
equations (in contrast to the ''fourth order'' metric formalism). Note that a linear Lagrangian $f(R) \sim R + 
2\Lambda$ corresponds to general relativity with a cosmological constant $\Lambda$. Note also that here we assume the 
nonlinear part of the function $f(R)$ to drive the cosmic acceleration so that there is no need for dark energy. 
One could attempt to eliminate also the need for dark matter with nonlinear gravity \cite{Borowiec:2006qr}. That is 
not our approach here, but our results could be relevant for such pursuits too.

The paper is organized as follows.
We generalize the cosmological equations in section \ref{cosmo} and apply them in 
section \ref{gene} to construct an $f(R)$GDM scenario, where anisotropically stressed dark matter 
has (up to a normalization) the $\Lambda$CDM (cold dark matter model with a cosmological constant) matter power 
spectrum, in a universe with its the background expansion identical to the corresponding $f(R)$CDM model. Therefore we can 
conclude in section \ref{conclu} that with specific assumptions about the properties of dark matter, the large-scale 
structure in these models can be consistent with observations. Appendix \ref{appe} contains the full linearized 
evolution equation.

\section{Cosmology with generalized dark matter}
\label{cosmo}

In the $\Lambda$CDM model the CDM component consists possibly of weakly 
interacting massive particles or something else which should be 
at cosmological scales very accurately approximated as an exactly 
pressureless and perfect fluid. These properties are deduced from 
the observations of large-scale clustering properties of matter, 
assuming the gravitational dynamics to be governed by GR. The so called 
hot dark matter scenario, as an example, is excluded because of the finite 
pressure in a hot component inhibits matter from clustering at subhorizon 
scales efficiently enough to explain the observed amount of structure at those scales
when the observations are interpreted within the framework of a cosmological
constant model \cite{Hu:1998tj}.

At present study we are focusing on an alternative gravity 
model featuring deviations from general relativity at the scales where observations require
to invoke dark matter. The question we then ask is not ''what 
gravity models are consistent with the CDM cosmology'', but 
rather {\it ''what properties of GDM are required
for cosmological viability of a given modified gravity model ?''}. 

To begin, we will derive an evolution equation for the inhomogeneities in a general fluid.
To characterize small perturbations about the background we can
without loss of generality adopt the longitudinal Newtonian gauge \cite{Mukhanov:1990me,Hu:1998tj}, which 
is defined by including the two gravitational potentials $\Psi$ and 
$\Phi$ to the Robertson-Walker line-element as
 \be \label{newtonian}
      ds^2 = a^2(\tau)[-(1+2\Psi)d\tau^2 + (1-2\Phi)g^{(3)}_{ij}dx^idx^j],
 \ee
where $g^{(3)}_{ij}$ is the spatial 3-metric which in a flat universe 
reduces to the Dirac delta function $g^{(3)}_{ij} = \delta_{ij}$. The components of 
the energy-momentum tensor for a general fluid including scalar perturbations 
can be written as
 \bea \label{fluid}
 T^0_{\phantom{0}0} & = & -\bar{\rho}\left(1+\delta^N\right), \\
 T^0_{\phantom{0}i} & = & -\left(\bar{\rho}+\bar{p}\right)v^N_{,i}, \\
 T^i_{\phantom{i}j} & = &
 \bar{\rho}(w + c^2\delta^N)\delta^i_j +
          \left(\nabla^i\nabla_j
         + \frac{1}{3}\delta^i_j\nabla^k\nabla_k\right)\Pi. 
 \eea
where an overbar means the background value, $\delta^N$ is the 
overdensity and $v^N$ the velocity potential evaluated in the Newtonian 
gauge. A nonzero equation of state $w = p/\rho$ would imply that the properties
of dark matter differ from CDM already in their effects to the overall
expansion of the universe. Then the pressure perturbation $\delta p$ could
differ from zero as well. Our description is not completely general as
we do not now allow entropic pressure, but 
assume\footnote{We assume this
for simplicity since if $w=0$ (which is satisfied by baryons and CDM), the entropic
pressure should identically vanish. On the other hand, a small sound speed
described by $c^2$ and negligible for the background, could be conceivable
for dark matter even when $w=0$ (this is indeed the standard description of baryons).} $\delta p = c^2 \delta \rho$,
where $c^2 \equiv \dot{p}/\dot{\rho}$ is the sound speed (an overdot denoting a derivative
with respect to the conformal time $\tau$). 
The anisotropic stress is constructed from the scalar potential $\Pi$ with the
aid of the covariant derivatives $\nabla_k$ of the metric $g^{(3)}_{ij}$.
The last term in Eq.(\ref{fluid}) can appear only at the perturbative level
in a Friedmann-Lemaitre universe. In the following, it will be useful to employ 
the gauge-invariant variable
\be \label{com}
\delta = \delta^N + 3H(1+w)\frac{v^N}{k},
\ee
This $\delta$ is equal to the fractional overdensity in the comoving 
gauge, where the velocity potential vanishes. On the other hand, it 
is proportional to the velocity potential in the uniform-density gauge \cite{Mukhanov:1990me}. 

Following the method of \cite{Koivisto:2005yc}, one may then derive a generalized
evolution equation for $\delta$ in the case where that energy-momentum
tensor is given by Eq.(\ref{fluid}). Here we report only the result, which is
\be \label{s_dddelta}
\ddot{\delta} =
D_1 H\dot{\delta} +
\left(D_2 H^2 + D_k k^2\right)\delta + 
P_1 H\dot{\Pi} +
P_2 H^2\Pi,
\ee
where the dimensionless coefficients are given explicitly in the Appendix \ref{appe}.
Eq.(\ref{s_dddelta}) determines completely the behaviour of cosmological fluctuations with
freedom to choose any $f$, $w$, $c^2$ or $\Pi$. Once the comoving matter
fluctuation $\delta$ is solved from this second order differential equation, the
corresponding matter variables in any other gauge, as well any metric perturbation, in
particular the potentials $\Psi$ and $\Phi$ in Eq.(\ref{newtonian}), are
uniquely fixed by the solution for $\delta$, and can be found as usually
by plugging this solution into the gauge transformation or constraint 
equations determining the relationships between the different variables. 
When $w=c^2=\Pi=0$, Eq.(\ref{s_dddelta}) reduces to the cases considered in
\cite{Koivisto:2005yc}. The term of main interest to us here is the gradient 
term Eq.(\ref{gradient}), whose effects 
easily spoil the agreement of these models with observations. This has 
been extensively discussed elsewhere \cite{Koivisto:2005yc,Koivisto:2006ie,Li:2006vi,Li:2006ag,Uddin:2007gj} 
and here we will just discuss the possibilities of alleviating the 
effects of this scale-dependent term. Let us though note that its presence stems from the 
new effective matter sources peculiar to the particular class of modified gravity models. It has been 
recently found that such a scale-dependent term is absent in various modified gravity theories within the metric formalism
\cite{Bean:2006up,Koivisto:2006xf,Koivisto:2006ai,Li:2007xn}.

\section{A specific example escaping the constraints} 
\label{gene}

One can notice that the dynamics of perturbations 
in a perfect fluid (with $\Pi=0$) can be scale-independent in the 
case that the sound speed of the fluid would obey 
\be \label{sound}
c^2 = -\frac{1}{6}\frac{d\log{F}}{d\log{a}},
\ee
where we have defined $F \equiv \partial f/\partial R$.
In the following we will however study an imperfect case with $\Pi \neq 0$.

We then use a specific expression for the shear stress potential of the cosmic fluid:
\be \label{shear}
\Pi = \frac{1}{1-3K/k^2}\left(\frac{\dot{F}}{4HF} + \frac{3}{2}c^2\right)\delta.
\ee
We have kept the prefactor $(1-3K/k^2)^{-1}$ here for the sake of generality, 
though it reduces to unity for scales smaller than the curvature radius of the universe, and 
in a flat universe it equals identically $1$. The first contribution in 
the parenthesis to $\Pi$ is there in order to eliminate the effective matter 
sources due to modified gravity, and the second in order to cancel the sources 
due to possible isotropic pressure in any kind of warm dark matter. What one then 
finds with this input is that Eq.(\ref{s_dddelta}) can be written in a simpler form 
\be \label{s_dddelta2}
\ddot{\delta} + \tilde{D}_1 H\dot{\delta} + \tilde{D}_2 H^2\delta = 0.
\ee
This verifies our claim that the effect of gradient can be cancelled by 
inherent properties of dark matter.

We can even consider the case when this happens while our dark matter fluid at 
the background level is completely pressureless as usual\footnote{Though our present 
approach is phenomenological and not aimed at explaining the origin
of dark matter stress $\Pi$, we might note that cosmic media satisfying effectively
$w=0$ but $\Pi \neq 0$ has indeed been considered \cite{Battye:2007aa}. 
(See also \cite{Koivisto:2005mm} for a recent discussion anisotropically stressed 
cosmological fluids and \cite{Hu:1998tj} for an extensive review of structure 
formation with generalized dark matter).}. Setting $w=0$ our result 
will further simplify to Eq.(\ref{s_dddelta3}) in Appendix \ref{appe}.
Thus we have shown that it is possible to avoid the tight 
constraints from structure formation, while keeping the background 
expansion exactly the same as in the CDM scenario. In fact, as the evolution in 
Eqs.(\ref{s_dddelta2},\ref{s_dddelta3}) 
is scale-invariant, the shape of the matter power spectrum is 
exactly the same for any choice of $w$, $c^2$, and in particular for any
function $f(R)$. Therefore, if the normalization is left arbitrary, no 
constraints at all arise from comparison with the shape of the observed  
matter power spectrum. However, the linear growth rate would be a potentially 
useful test for these models.

To illustrate this, we will study the toy model $f \sim R^n$. This model 
allows an analytical treatment and its predictions have been compared with the
data on the late time cosmological 
expansion\cite{Capozziello:2004vh,Allemandi:2004ca,Borowiec:2006hk,Borowiec:2006qr}. In fact, the 
background is simply described by an effective equation of state $w_{eff} = -1 + (1+w)/n$ when $K=0$, so 
that then
\bea \label{r_hub}
H^2 & = & \frac{4n^2}{\left(3\left(1+w\right)-2n\right)^2}\frac{1}{\tau^2}, \\
\dot{H} & = & \left(1-\frac{3\left(1+w\right)}{2n}\right)H^2, \\ 
\ddot{H} & = & \frac{1}{2}\left(2-\frac{3\left(1+w\right)}{n}\right)^2 H^3.
\eea
It is also easy to see also that
\bea \label{r_eff}
\dot{F} & = & \frac{3\left(1+w\right)\left(1-n\right)}{n}HF, \\
\ddot{F} & = & 3\left[\frac{1}{3} - w + c^2 
+ \left(1+w\right)\left(\frac{1}{2n}-1\right)\right]H\dot{F}.
\eea
Plugging then the expressions (\ref{r_hub}) and (\ref{r_eff}) in Eq.(\ref{s_dddelta2}) one 
finds that, keeping $w$ constant for simplicity, 
 \be
\ddot{\delta}   =   
\frac{3\left(1+w\right)  - n\left(1+9w\right)}{2n-3\left(1+w\right)}\frac{\dot{\delta}}{\tau} +
\frac{6\left(n-2\right)\left(1+w\right)}{2n-3\left(1+w\right)}\frac{\delta}{\tau^2}.
 \ee
This admits two power-law solutions, the other one corresponding to a 
decaying and the other to a growing mode. Setting further $w=0$ we find 
that the latter is characterized by the growth rate 
\be \label{r}
r \equiv \frac{d \log{\delta}}{d \log{a}} =  \frac{|n-|12-7n||}{4n}.
\ee
There is an observational estimate \cite{Verde:2001sf}
for this $r$ at the redhift $z = 
1/a-1=0.15$ from the 2dFGRS data \cite{Colless:2001gk}, implying that $r = 0.51 \pm 0.11$. For the $f(R) 
\sim R^n$ model considered here, an analysis combining the supernova 
and baryon oscillation scale constraints shows that the background data
prefers the values of the exponent $n = 2.6 \pm 0.3$ \cite{Borowiec:2006hk}. 
This corresponds to about $0.20 < r < 0.47$, thus agreeing with the
2dFGRS constraints. Note that the linear evolution for this same model, but without 
making the assumption Eq.(\ref{shear}), is in gross disagreement with the 
data\cite{Koivisto:2005yc}. One may also quantify the linear behaviour
utilizing the growth index $\gamma$ defined via \cite{Huterer:2006mv}
\be
g(a) \equiv \delta/a = e^{\int_0^a 
d\log{a}\left(\Omega_m(a)^\gamma-1\right)},
\ee
This single-parameter minimalist characterization of modified gravity effects
turns out to depend both on the matter density and about the exponent $n$, whereas 
the result for the growth rate $r$ was independent of the amount of matter. 
We find a simple relation, $\gamma = \log{r}/\log{\Omega_m}$. As the background 
data constrains quite tightly $\Omega_m \approx 0.3$ \cite{Borowiec:2006hk}, we have
$0.62 < \gamma < 1.34$. There are other useful probes of modified gravity 
effects \cite{Zhang:2007nk}, but their detailed analysis is left for further studies.

\section{Conclusions}
\label{conclu}

Inspired by the the theoretical developments \cite{Sotiriou:2006qn}, 
the success of the Palatini-$f(R)$ gravities within the Solar system \cite{Kainulainen:2006wz}
and the recent resolutions of their physical implications \cite{Kainulainen:2007bt}, we 
reconsidered the cosmological viability of these models in view of their predictions for the large-scale structure. 

Previous investigations on the subject had established that the constraints from 
matter power spectrum allow only models which are practically indistinguishable (by e.g. their background expansion)
from the $\Lambda$CDM model \cite{Koivisto:2005yc,Koivisto:2006ie,Li:2006vi,Li:2006ag}. In the present 
paper we examined the robustness of those conclusions to the variations of the dark matter scenario. 
We pointed out that the extremely tight constraints from structure
formation are valid only for the $f(R)$CDM models, i.e. when the universe energy density 
is assumed to be dominated by pressureless and perfect matter. 
The instabilities occurring then due to modified gravity effects, may be less severe or even absent
in $f(R)$GDM scenarios, i.e. when dark matter is allowed to have inherent stresses. The cosmological 
bounds on these $f(R)$ models could thereby be drastically loosened. 

This demonstrates that a revision of the nature dark energy could also change our view of dark matter. 

\acknowledgments

The author is grateful to the Magnus Ehrnrooth Foundation and the 
Finnish Cultural Foundation for their support. Marie Curie Research and 
Training Network ''UniverseNet'' (HPRN-CT-2006-035863) is acknowledged.

\begin{widetext}

\appendix 

\section{Evolution equation}
\label{appe}

The equation (\ref{s_dddelta}) was written as
\be 
\ddot{\delta} =
D_1 H\dot{\delta} +
\left(D_2 H^2 + D_k k^2\right) \delta +
P_1 H\dot{\Pi} +
P_2 H^2\Pi.
\ee
Defining $F \equiv \partial f/\partial R$, we can then write the dimensionless coefficients as the following:
\be
D_1 =  \frac{-2FH\left(FH^2(1+3c^2-6w) + \ddot{F} \right) + 2\dot{F}^2H + \dot{F}F\left(2\dot{H}-H^2(1+3c^2-6w)\right)
                                    }{F H\left( \dot{F} + 2FH \right)},
\ee
\bea
FH^2\left(\dot{F} + 2FH \right)D_2  & = &   
2FH\left[ FH\left(12H^3(w-c^2)+\dot{H}H(3c^2+3w-2) + \ddot{H}\right) +  \ddot{F}\left(\dot{H}-H^2(1-3w)\right)\right] \\ \nonumber & + &
\dot{F}F\left(-2\dot{H}^2+H\ddot{H}+3H^2(\dot{H}-4H^2)(c^2-w)\right) - 2\dot{F}^2H\left(\dot{H}-H^2(1-3w)\right)
\eea     
\be \label{gradient}
D_k = \frac{-\left( \dot{F} + 6c^2FH \right)}{3\left( \dot{F} + 2FH \right)},
\ee
\be
P_1 = 2\left(\frac{3K}{k^2}-1\right),
\ee
and
\be
P_2 =    \frac{2\left(k^2 - 3K \right)
      \left[ 6\dot{F}^2 +
        9\dot{F}FH\left(c^2-w\right)  -
        2F\left(3\ddot{F} +
           F( 6\dot{H} + 9(c^2-w) H^2 - k^2 )\right)  \right] }{3FH
      \left( \dot{F} + 2FH \right)k^2}.
\ee
In the case that Eq.(\ref{shear}) holds and $w=0$, the evolution equation reduces to 
\bea \label{s_dddelta3}
\ddot{\delta} & = &
\frac{1}{2F^2H^2\left( \dot{F} + 2FH \right)}
\Bigg\{ FH\Big[-4FH\left(FH^2+\ddot{F} \right) + 3\dot{F}^2H +
      4\dot{F}F\left(\dot{H} - H^2\right)  \Big]\dot{\delta}
  \\ \nonumber &  + &  
\Big[
2F^2H\left( 2FH( \ddot{H} - 2\dot{H}H)  +
         \ddot{F}(2\dot{H} - 3H^2)
        \right) + 3\dot{F}^3H^2 - 3\dot{F}^2FH( \dot{H} - 2H^2) \\ \nonumber
& + &    \dot{F}F\left(-4\dot{H}^2F +
         2\ddot{H}FH -
         (3\ddot{F} + 2\dot{H}F) H^2\right)  
\Big]\delta\Bigg\}.
\eea

\end{widetext}

\bibliography{refs}

\begin{thebibliography}{49}
\expandafter\ifx\csname natexlab\endcsname\relax\def\natexlab#1{#1}\fi
\expandafter\ifx\csname bibnamefont\endcsname\relax
  \def\bibnamefont#1{#1}\fi
\expandafter\ifx\csname bibfnamefont\endcsname\relax
  \def\bibfnamefont#1{#1}\fi
\expandafter\ifx\csname citenamefont\endcsname\relax
  \def\citenamefont#1{#1}\fi
\expandafter\ifx\csname url\endcsname\relax
  \def\url#1{\texttt{#1}}\fi
\expandafter\ifx\csname urlprefix\endcsname\relax\def\urlprefix{URL }\fi
\providecommand{\bibinfo}[2]{#2}
\providecommand{\eprint}[2][]{\url{#2}}

\bibitem[{\citenamefont{Nojiri and Odintsov}(2007)}]{Nojiri:2006ri}
\bibinfo{author}{\bibfnamefont{S.}~\bibnamefont{Nojiri}} \bibnamefont{and}
  \bibinfo{author}{\bibfnamefont{S.~D.} \bibnamefont{Odintsov}},
  \bibinfo{journal}{Int. J. Geom. Meth. Mod. Phys.}
  \textbf{\bibinfo{volume}{4}}, \bibinfo{pages}{115} (\bibinfo{year}{2007}),
  \eprint{hep-th/0601213}.

\bibitem[{\citenamefont{Carroll et~al.}(2004)\citenamefont{Carroll, Duvvuri,
  Trodden, and Turner}}]{Carroll:2003wy}
\bibinfo{author}{\bibfnamefont{S.~M.} \bibnamefont{Carroll}},
  \bibinfo{author}{\bibfnamefont{V.}~\bibnamefont{Duvvuri}},
  \bibinfo{author}{\bibfnamefont{M.}~\bibnamefont{Trodden}}, \bibnamefont{and}
  \bibinfo{author}{\bibfnamefont{M.~S.} \bibnamefont{Turner}},
  \bibinfo{journal}{Phys. Rev.} \textbf{\bibinfo{volume}{D70}},
  \bibinfo{pages}{043528} (\bibinfo{year}{2004}), \eprint{astro-ph/0306438}.

\bibitem[{\citenamefont{Nojiri and
  Odintsov}(2003{\natexlab{a}})}]{Nojiri:2003rz}
\bibinfo{author}{\bibfnamefont{S.}~\bibnamefont{Nojiri}} \bibnamefont{and}
  \bibinfo{author}{\bibfnamefont{S.~D.} \bibnamefont{Odintsov}},
  \bibinfo{journal}{Phys. Lett.} \textbf{\bibinfo{volume}{B576}},
  \bibinfo{pages}{5} (\bibinfo{year}{2003}{\natexlab{a}}),
  \eprint{hep-th/0307071}.

\bibitem[{\citenamefont{Nojiri and
  Odintsov}(2003{\natexlab{b}})}]{Nojiri:2003ft}
\bibinfo{author}{\bibfnamefont{S.}~\bibnamefont{Nojiri}} \bibnamefont{and}
  \bibinfo{author}{\bibfnamefont{S.~D.} \bibnamefont{Odintsov}},
  \bibinfo{journal}{Phys. Rev.} \textbf{\bibinfo{volume}{D68}},
  \bibinfo{pages}{123512} (\bibinfo{year}{2003}{\natexlab{b}}),
  \eprint{hep-th/0307288}.

\bibitem[{\citenamefont{Faraoni and Nadeau}(2007)}]{Faraoni:2006fx}
\bibinfo{author}{\bibfnamefont{V.}~\bibnamefont{Faraoni}} \bibnamefont{and}
  \bibinfo{author}{\bibfnamefont{S.}~\bibnamefont{Nadeau}},
  \bibinfo{journal}{Phys. Rev.} \textbf{\bibinfo{volume}{D75}},
  \bibinfo{pages}{023501} (\bibinfo{year}{2007}), \eprint{gr-qc/0612075}.

\bibitem[{\citenamefont{Faraoni}(2005)}]{Faraoni:2005ie}
\bibinfo{author}{\bibfnamefont{V.}~\bibnamefont{Faraoni}},
  \bibinfo{journal}{Phys. Rev.} \textbf{\bibinfo{volume}{D72}},
  \bibinfo{pages}{061501} (\bibinfo{year}{2005}), \eprint{gr-qc/0509008}.

\bibitem[{\citenamefont{Nojiri and Odintsov}(2006)}]{Nojiri:2006gh}
\bibinfo{author}{\bibfnamefont{S.}~\bibnamefont{Nojiri}} \bibnamefont{and}
  \bibinfo{author}{\bibfnamefont{S.~D.} \bibnamefont{Odintsov}},
  \bibinfo{journal}{Phys. Rev.} \textbf{\bibinfo{volume}{D74}},
  \bibinfo{pages}{086005} (\bibinfo{year}{2006}), \eprint{hep-th/0608008}.

\bibitem[{\citenamefont{Capozziello
  et~al.}(2006{\natexlab{a}})\citenamefont{Capozziello, Nojiri, Odintsov, and
  Troisi}}]{Capozziello:2006dj}
\bibinfo{author}{\bibfnamefont{S.}~\bibnamefont{Capozziello}},
  \bibinfo{author}{\bibfnamefont{S.}~\bibnamefont{Nojiri}},
  \bibinfo{author}{\bibfnamefont{S.~D.} \bibnamefont{Odintsov}},
  \bibnamefont{and} \bibinfo{author}{\bibfnamefont{A.}~\bibnamefont{Troisi}},
  \bibinfo{journal}{Phys. Lett.} \textbf{\bibinfo{volume}{B639}},
  \bibinfo{pages}{135} (\bibinfo{year}{2006}{\natexlab{a}}),
  \eprint{astro-ph/0604431}.

\bibitem[{\citenamefont{Chiba}(2003)}]{Chiba:2003ir}
\bibinfo{author}{\bibfnamefont{T.}~\bibnamefont{Chiba}},
  \bibinfo{journal}{Phys. Lett.} \textbf{\bibinfo{volume}{B575}},
  \bibinfo{pages}{1} (\bibinfo{year}{2003}), \eprint{astro-ph/0307338}.

\bibitem[{\citenamefont{Erickcek et~al.}(2006)\citenamefont{Erickcek, Smith,
  and Kamionkowski}}]{Erickcek:2006vf}
\bibinfo{author}{\bibfnamefont{A.~L.} \bibnamefont{Erickcek}},
  \bibinfo{author}{\bibfnamefont{T.~L.} \bibnamefont{Smith}}, \bibnamefont{and}
  \bibinfo{author}{\bibfnamefont{M.}~\bibnamefont{Kamionkowski}},
  \bibinfo{journal}{Phys. Rev.} \textbf{\bibinfo{volume}{D74}},
  \bibinfo{pages}{121501} (\bibinfo{year}{2006}), \eprint{astro-ph/0610483}.

\bibitem[{\citenamefont{Kainulainen et~al.}(2007)\citenamefont{Kainulainen,
  Piilonen, Reijonen, and Sunhede}}]{Kainulainen:2007bt}
\bibinfo{author}{\bibfnamefont{K.}~\bibnamefont{Kainulainen}},
  \bibinfo{author}{\bibfnamefont{J.}~\bibnamefont{Piilonen}},
  \bibinfo{author}{\bibfnamefont{V.}~\bibnamefont{Reijonen}}, \bibnamefont{and}
  \bibinfo{author}{\bibfnamefont{D.}~\bibnamefont{Sunhede}}
  (\bibinfo{year}{2007}), \eprint{arXiv:0704.2729 [gr-qc]}.

\bibitem[{\citenamefont{Hu and Sawicki}(2007)}]{Hu:2007nk}
\bibinfo{author}{\bibfnamefont{W.}~\bibnamefont{Hu}} \bibnamefont{and}
  \bibinfo{author}{\bibfnamefont{I.}~\bibnamefont{Sawicki}}
  (\bibinfo{year}{2007}), \eprint{arXiv:0705.1158 [astro-ph]}.

\bibitem[{\citenamefont{Ferraris et~al.}(1992)\citenamefont{Ferraris,
  Francaviglia, and Volovich}}]{Ferraris:1992dx}
\bibinfo{author}{\bibfnamefont{M.}~\bibnamefont{Ferraris}},
  \bibinfo{author}{\bibfnamefont{M.}~\bibnamefont{Francaviglia}},
  \bibnamefont{and} \bibinfo{author}{\bibfnamefont{I.}~\bibnamefont{Volovich}}
  (\bibinfo{year}{1992}), \eprint{gr-qc/9303007}.

\bibitem[{\citenamefont{Vollick}(2003)}]{Vollick:2003aw}
\bibinfo{author}{\bibfnamefont{D.~N.} \bibnamefont{Vollick}},
  \bibinfo{journal}{Phys. Rev.} \textbf{\bibinfo{volume}{D68}},
  \bibinfo{pages}{063510} (\bibinfo{year}{2003}), \eprint{astro-ph/0306630}.

\bibitem[{\citenamefont{Flanagan}(2003)}]{Flanagan:2003iw}
\bibinfo{author}{\bibfnamefont{E.~E.} \bibnamefont{Flanagan}},
  \bibinfo{journal}{Class. Quant. Grav.} \textbf{\bibinfo{volume}{21}},
  \bibinfo{pages}{417} (\bibinfo{year}{2003}), \eprint{gr-qc/0309015}.

\bibitem[{\citenamefont{Meng and Wang}(2004)}]{Meng:2003en}
\bibinfo{author}{\bibfnamefont{X.-H.} \bibnamefont{Meng}} \bibnamefont{and}
  \bibinfo{author}{\bibfnamefont{P.}~\bibnamefont{Wang}},
  \bibinfo{journal}{Phys. Lett.} \textbf{\bibinfo{volume}{B584}},
  \bibinfo{pages}{1} (\bibinfo{year}{2004}), \eprint{hep-th/0309062}.

\bibitem[{\citenamefont{Allemandi et~al.}(2005)\citenamefont{Allemandi,
  Borowiec, Francaviglia, and Odintsov}}]{Allemandi:2005qs}
\bibinfo{author}{\bibfnamefont{G.}~\bibnamefont{Allemandi}},
  \bibinfo{author}{\bibfnamefont{A.}~\bibnamefont{Borowiec}},
  \bibinfo{author}{\bibfnamefont{M.}~\bibnamefont{Francaviglia}},
  \bibnamefont{and} \bibinfo{author}{\bibfnamefont{S.~D.}
  \bibnamefont{Odintsov}}, \bibinfo{journal}{Phys. Rev.}
  \textbf{\bibinfo{volume}{D72}}, \bibinfo{pages}{063505}
  (\bibinfo{year}{2005}), \eprint{gr-qc/0504057}.

\bibitem[{\citenamefont{Poplawski}(2006)}]{Poplawski:2006kv}
\bibinfo{author}{\bibfnamefont{N.~J.} \bibnamefont{Poplawski}},
  \bibinfo{journal}{Phys. Rev.} \textbf{\bibinfo{volume}{D74}},
  \bibinfo{pages}{084032} (\bibinfo{year}{2006}), \eprint{gr-qc/0607124}.

\bibitem[{\citenamefont{Kainulainen et~al.}(2006)\citenamefont{Kainulainen,
  Reijonen, and Sunhede}}]{Kainulainen:2006wz}
\bibinfo{author}{\bibfnamefont{K.}~\bibnamefont{Kainulainen}},
  \bibinfo{author}{\bibfnamefont{V.}~\bibnamefont{Reijonen}}, \bibnamefont{and}
  \bibinfo{author}{\bibfnamefont{D.}~\bibnamefont{Sunhede}}
  (\bibinfo{year}{2006}), \eprint{gr-qc/0611132}.

\bibitem[{\citenamefont{Flanagan}(2004)}]{Flanagan:2003rb}
\bibinfo{author}{\bibfnamefont{E.~E.} \bibnamefont{Flanagan}},
  \bibinfo{journal}{Phys. Rev. Lett.} \textbf{\bibinfo{volume}{92}},
  \bibinfo{pages}{071101} (\bibinfo{year}{2004}), \eprint{astro-ph/0308111}.

\bibitem[{\citenamefont{Vollick}(2004)}]{Vollick:2003ic}
\bibinfo{author}{\bibfnamefont{D.~N.} \bibnamefont{Vollick}},
  \bibinfo{journal}{Class. Quant. Grav.} \textbf{\bibinfo{volume}{21}},
  \bibinfo{pages}{3813} (\bibinfo{year}{2004}), \eprint{gr-qc/0312041}.

\bibitem[{\citenamefont{Olmo}(2007)}]{Olmo:2006zu}
\bibinfo{author}{\bibfnamefont{G.~J.} \bibnamefont{Olmo}},
  \bibinfo{journal}{Phys. Rev. Lett.} \textbf{\bibinfo{volume}{98}},
  \bibinfo{pages}{061101} (\bibinfo{year}{2007}), \eprint{gr-qc/0612002}.

\bibitem[{\citenamefont{Koivisto}(2006{\natexlab{a}})}]{Koivisto:2005yk}
\bibinfo{author}{\bibfnamefont{T.}~\bibnamefont{Koivisto}},
  \bibinfo{journal}{Class. Quant. Grav.} \textbf{\bibinfo{volume}{23}},
  \bibinfo{pages}{4289} (\bibinfo{year}{2006}{\natexlab{a}}),
  \eprint{gr-qc/0505128}.

\bibitem[{\citenamefont{Barausse et~al.}(2007)\citenamefont{Barausse, Sotiriou,
  and Miller}}]{Barausse:2007pn}
\bibinfo{author}{\bibfnamefont{E.}~\bibnamefont{Barausse}},
  \bibinfo{author}{\bibfnamefont{T.~P.} \bibnamefont{Sotiriou}},
  \bibnamefont{and} \bibinfo{author}{\bibfnamefont{J.~C.} \bibnamefont{Miller}}
  (\bibinfo{year}{2007}), \eprint{gr-qc/0703132}.

\bibitem[{\citenamefont{Sotiriou and Liberati}(2007)}]{Sotiriou:2006qn}
\bibinfo{author}{\bibfnamefont{T.~P.} \bibnamefont{Sotiriou}} \bibnamefont{and}
  \bibinfo{author}{\bibfnamefont{S.}~\bibnamefont{Liberati}},
  \bibinfo{journal}{Annals Phys.} \textbf{\bibinfo{volume}{322}},
  \bibinfo{pages}{935} (\bibinfo{year}{2007}), \eprint{gr-qc/0604006}.

\bibitem[{\citenamefont{Capozziello
  et~al.}(2006{\natexlab{b}})\citenamefont{Capozziello, Cardone, and
  Francaviglia}}]{Capozziello:2004vh}
\bibinfo{author}{\bibfnamefont{S.}~\bibnamefont{Capozziello}},
  \bibinfo{author}{\bibfnamefont{V.~F.} \bibnamefont{Cardone}},
  \bibnamefont{and}
  \bibinfo{author}{\bibfnamefont{M.}~\bibnamefont{Francaviglia}},
  \bibinfo{journal}{Gen. Rel. Grav.} \textbf{\bibinfo{volume}{38}},
  \bibinfo{pages}{711} (\bibinfo{year}{2006}{\natexlab{b}}),
  \eprint{astro-ph/0410135}.

\bibitem[{\citenamefont{Amarzguioui et~al.}(2006)\citenamefont{Amarzguioui,
  Elgaroy, Mota, and Multamaki}}]{Amarzguioui:2005zq}
\bibinfo{author}{\bibfnamefont{M.}~\bibnamefont{Amarzguioui}},
  \bibinfo{author}{\bibfnamefont{O.}~\bibnamefont{Elgaroy}},
  \bibinfo{author}{\bibfnamefont{D.~F.} \bibnamefont{Mota}}, \bibnamefont{and}
  \bibinfo{author}{\bibfnamefont{T.}~\bibnamefont{Multamaki}},
  \bibinfo{journal}{Astron. Astrophys.} \textbf{\bibinfo{volume}{454}},
  \bibinfo{pages}{707} (\bibinfo{year}{2006}), \eprint{astro-ph/0510519}.

\bibitem[{\citenamefont{Borowiec et~al.}(2006)\citenamefont{Borowiec,
  Godlowski, and Szydlowski}}]{Borowiec:2006hk}
\bibinfo{author}{\bibfnamefont{A.}~\bibnamefont{Borowiec}},
  \bibinfo{author}{\bibfnamefont{W.}~\bibnamefont{Godlowski}},
  \bibnamefont{and}
  \bibinfo{author}{\bibfnamefont{M.}~\bibnamefont{Szydlowski}},
  \bibinfo{journal}{Phys. Rev.} \textbf{\bibinfo{volume}{D74}},
  \bibinfo{pages}{043502} (\bibinfo{year}{2006}), \eprint{astro-ph/0602526}.

\bibitem[{\citenamefont{Movahed et~al.}(2007)\citenamefont{Movahed, Baghram,
  and Rahvar}}]{Movahed:2007cs}
\bibinfo{author}{\bibfnamefont{M.~S.} \bibnamefont{Movahed}},
  \bibinfo{author}{\bibfnamefont{S.}~\bibnamefont{Baghram}}, \bibnamefont{and}
  \bibinfo{author}{\bibfnamefont{S.}~\bibnamefont{Rahvar}}
  (\bibinfo{year}{2007}), \eprint{arXiv:0705.0889 [astro-ph]}.

\bibitem[{\citenamefont{Fay et~al.}(2007)\citenamefont{Fay, Tavakol, and
  Tsujikawa}}]{Fay:2007gg}
\bibinfo{author}{\bibfnamefont{S.}~\bibnamefont{Fay}},
  \bibinfo{author}{\bibfnamefont{R.}~\bibnamefont{Tavakol}}, \bibnamefont{and}
  \bibinfo{author}{\bibfnamefont{S.}~\bibnamefont{Tsujikawa}}
  (\bibinfo{year}{2007}), \eprint{astro-ph/0701479}.

\bibitem[{\citenamefont{Koivisto and Kurki-Suonio}(2006)}]{Koivisto:2005yc}
\bibinfo{author}{\bibfnamefont{T.}~\bibnamefont{Koivisto}} \bibnamefont{and}
  \bibinfo{author}{\bibfnamefont{H.}~\bibnamefont{Kurki-Suonio}},
  \bibinfo{journal}{Class. Quant. Grav.} \textbf{\bibinfo{volume}{23}},
  \bibinfo{pages}{2355} (\bibinfo{year}{2006}), \eprint{astro-ph/0509422}.

\bibitem[{\citenamefont{Koivisto}(2006{\natexlab{b}})}]{Koivisto:2006ie}
\bibinfo{author}{\bibfnamefont{T.}~\bibnamefont{Koivisto}},
  \bibinfo{journal}{Phys. Rev.} \textbf{\bibinfo{volume}{D73}},
  \bibinfo{pages}{083517} (\bibinfo{year}{2006}{\natexlab{b}}),
  \eprint{astro-ph/0602031}.

\bibitem[{\citenamefont{Li and Chu}(2006)}]{Li:2006vi}
\bibinfo{author}{\bibfnamefont{B.}~\bibnamefont{Li}} \bibnamefont{and}
  \bibinfo{author}{\bibfnamefont{M.~C.} \bibnamefont{Chu}},
  \bibinfo{journal}{Phys. Rev.} \textbf{\bibinfo{volume}{D74}},
  \bibinfo{pages}{104010} (\bibinfo{year}{2006}), \eprint{astro-ph/0610486}.

\bibitem[{\citenamefont{Li et~al.}(2006)\citenamefont{Li, Chan, and
  Chu}}]{Li:2006ag}
\bibinfo{author}{\bibfnamefont{B.}~\bibnamefont{Li}},
  \bibinfo{author}{\bibfnamefont{K.~C.} \bibnamefont{Chan}}, \bibnamefont{and}
  \bibinfo{author}{\bibfnamefont{M.~C.} \bibnamefont{Chu}}
  (\bibinfo{year}{2006}), \eprint{astro-ph/0610794}.

\bibitem[{\citenamefont{Borowiec et~al.}(2007)\citenamefont{Borowiec,
  Godlowski, and Szydlowski}}]{Borowiec:2006qr}
\bibinfo{author}{\bibfnamefont{A.}~\bibnamefont{Borowiec}},
  \bibinfo{author}{\bibfnamefont{W.}~\bibnamefont{Godlowski}},
  \bibnamefont{and}
  \bibinfo{author}{\bibfnamefont{M.}~\bibnamefont{Szydlowski}},
  \bibinfo{journal}{Int. J. Geom. Meth. Mod. Phys.}
  \textbf{\bibinfo{volume}{4}}, \bibinfo{pages}{183} (\bibinfo{year}{2007}),
  \eprint{astro-ph/0607639}.

\bibitem[{\citenamefont{Hu and Eisenstein}(1999)}]{Hu:1998tj}
\bibinfo{author}{\bibfnamefont{W.}~\bibnamefont{Hu}} \bibnamefont{and}
  \bibinfo{author}{\bibfnamefont{D.~J.} \bibnamefont{Eisenstein}},
  \bibinfo{journal}{Phys. Rev.} \textbf{\bibinfo{volume}{D59}},
  \bibinfo{pages}{083509} (\bibinfo{year}{1999}), \eprint{astro-ph/9809368}.

\bibitem[{\citenamefont{Mukhanov et~al.}(1992)\citenamefont{Mukhanov, Feldman,
  and Brandenberger}}]{Mukhanov:1990me}
\bibinfo{author}{\bibfnamefont{V.~F.} \bibnamefont{Mukhanov}},
  \bibinfo{author}{\bibfnamefont{H.~A.} \bibnamefont{Feldman}},
  \bibnamefont{and} \bibinfo{author}{\bibfnamefont{R.~H.}
  \bibnamefont{Brandenberger}}, \bibinfo{journal}{Phys. Rept.}
  \textbf{\bibinfo{volume}{215}}, \bibinfo{pages}{203} (\bibinfo{year}{1992}).

\bibitem[{\citenamefont{Uddin et~al.}(2007)\citenamefont{Uddin, Lidsey, and
  Tavakol}}]{Uddin:2007gj}
\bibinfo{author}{\bibfnamefont{K.}~\bibnamefont{Uddin}},
  \bibinfo{author}{\bibfnamefont{J.~E.} \bibnamefont{Lidsey}},
  \bibnamefont{and} \bibinfo{author}{\bibfnamefont{R.}~\bibnamefont{Tavakol}}
  (\bibinfo{year}{2007}), \eprint{arXiv:0705.0232 [gr-qc]}.

\bibitem[{\citenamefont{Bean et~al.}(2006)\citenamefont{Bean, Bernat, Pogosian,
  Silvestri, and Trodden}}]{Bean:2006up}
\bibinfo{author}{\bibfnamefont{R.}~\bibnamefont{Bean}},
  \bibinfo{author}{\bibfnamefont{D.}~\bibnamefont{Bernat}},
  \bibinfo{author}{\bibfnamefont{L.}~\bibnamefont{Pogosian}},
  \bibinfo{author}{\bibfnamefont{A.}~\bibnamefont{Silvestri}},
  \bibnamefont{and} \bibinfo{author}{\bibfnamefont{M.}~\bibnamefont{Trodden}}
  (\bibinfo{year}{2006}), \eprint{astro-ph/0611321}.

\bibitem[{\citenamefont{Koivisto and
  Mota}(2007{\natexlab{a}})}]{Koivisto:2006xf}
\bibinfo{author}{\bibfnamefont{T.}~\bibnamefont{Koivisto}} \bibnamefont{and}
  \bibinfo{author}{\bibfnamefont{D.~F.} \bibnamefont{Mota}},
  \bibinfo{journal}{Phys. Lett.} \textbf{\bibinfo{volume}{B644}},
  \bibinfo{pages}{104} (\bibinfo{year}{2007}{\natexlab{a}}),
  \eprint{astro-ph/0606078}.

\bibitem[{\citenamefont{Koivisto and
  Mota}(2007{\natexlab{b}})}]{Koivisto:2006ai}
\bibinfo{author}{\bibfnamefont{T.}~\bibnamefont{Koivisto}} \bibnamefont{and}
  \bibinfo{author}{\bibfnamefont{D.~F.} \bibnamefont{Mota}},
  \bibinfo{journal}{Phys. Rev.} \textbf{\bibinfo{volume}{D75}},
  \bibinfo{pages}{023518} (\bibinfo{year}{2007}{\natexlab{b}}),
  \eprint{hep-th/0609155}.

\bibitem[{\citenamefont{Li and Barrow}(2007)}]{Li:2007xn}
\bibinfo{author}{\bibfnamefont{B.}~\bibnamefont{Li}} \bibnamefont{and}
  \bibinfo{author}{\bibfnamefont{J.~D.} \bibnamefont{Barrow}}
  (\bibinfo{year}{2007}), \eprint{gr-qc/0701111}.

\bibitem[{\citenamefont{Battye and Moss}(2007)}]{Battye:2007aa}
\bibinfo{author}{\bibfnamefont{R.~A.} \bibnamefont{Battye}} \bibnamefont{and}
  \bibinfo{author}{\bibfnamefont{A.}~\bibnamefont{Moss}}
  (\bibinfo{year}{2007}), \eprint{astro-ph/0703744}.

\bibitem[{\citenamefont{Koivisto and Mota}(2006)}]{Koivisto:2005mm}
\bibinfo{author}{\bibfnamefont{T.}~\bibnamefont{Koivisto}} \bibnamefont{and}
  \bibinfo{author}{\bibfnamefont{D.~F.} \bibnamefont{Mota}},
  \bibinfo{journal}{Phys. Rev.} \textbf{\bibinfo{volume}{D73}},
  \bibinfo{pages}{083502} (\bibinfo{year}{2006}), \eprint{astro-ph/0512135}.

\bibitem[{\citenamefont{Allemandi et~al.}(2004)\citenamefont{Allemandi,
  Borowiec, and Francaviglia}}]{Allemandi:2004ca}
\bibinfo{author}{\bibfnamefont{G.}~\bibnamefont{Allemandi}},
  \bibinfo{author}{\bibfnamefont{A.}~\bibnamefont{Borowiec}}, \bibnamefont{and}
  \bibinfo{author}{\bibfnamefont{M.}~\bibnamefont{Francaviglia}},
  \bibinfo{journal}{Phys. Rev.} \textbf{\bibinfo{volume}{D70}},
  \bibinfo{pages}{043524} (\bibinfo{year}{2004}), \eprint{hep-th/0403264}.

\bibitem[{\citenamefont{Verde et~al.}(2002)}]{Verde:2001sf}
\bibinfo{author}{\bibfnamefont{L.}~\bibnamefont{Verde}} \bibnamefont{et~al.},
  \bibinfo{journal}{Mon. Not. Roy. Astron. Soc.}
  \textbf{\bibinfo{volume}{335}}, \bibinfo{pages}{432} (\bibinfo{year}{2002}),
  \eprint{astro-ph/0112161}.

\bibitem[{\citenamefont{Colless et~al.}(2001)}]{Colless:2001gk}
\bibinfo{author}{\bibfnamefont{M.}~\bibnamefont{Colless}} \bibnamefont{et~al.}
  (\bibinfo{collaboration}{The 2DFGRS}), \bibinfo{journal}{Mon. Not. Roy.
  Astron. Soc.} \textbf{\bibinfo{volume}{328}}, \bibinfo{pages}{1039}
  (\bibinfo{year}{2001}), \eprint{astro-ph/0106498}.

\bibitem[{\citenamefont{Huterer and Linder}(2007)}]{Huterer:2006mv}
\bibinfo{author}{\bibfnamefont{D.}~\bibnamefont{Huterer}} \bibnamefont{and}
  \bibinfo{author}{\bibfnamefont{E.~V.} \bibnamefont{Linder}},
  \bibinfo{journal}{Phys. Rev.} \textbf{\bibinfo{volume}{D75}},
  \bibinfo{pages}{023519} (\bibinfo{year}{2007}), \eprint{astro-ph/0608681}.

\bibitem[{\citenamefont{Zhang et~al.}(2007)\citenamefont{Zhang, Liguori, Bean,
  and Dodelson}}]{Zhang:2007nk}
\bibinfo{author}{\bibfnamefont{P.}~\bibnamefont{Zhang}},
  \bibinfo{author}{\bibfnamefont{M.}~\bibnamefont{Liguori}},
  \bibinfo{author}{\bibfnamefont{R.}~\bibnamefont{Bean}}, \bibnamefont{and}
  \bibinfo{author}{\bibfnamefont{S.}~\bibnamefont{Dodelson}}
  (\bibinfo{year}{2007}), \eprint{arXiv:0704.1932 [astro-ph]}.

\end{thebibliography}
\end{document}